# BUFFERED ELECTROPOLISHING --- A NEW WAY FOR ACHIEVING EXTREMELY SMOOTH SURFACE FINISH ON NB SRF CAVITIES TO BE USED IN PARTICLE ACCELERATORS*

A.T. Wu[#], R.A. Rimmer, X.Y. Lu[1], F. Eozenou[2], L. Lin[1], G. Ciovati, J. Mammosser[3], S. Jin[1], E.D. Wang[1], H. Tian, J. Williams, R. Manus, C. Reece, L. Phillips, and W. Sommer
Thomas Jefferson National Accelerator Facility, 12000 Jefferson Avenue, Newport News, VA 23606, USA
[1] SRF Group, Institute of Heavy Ion Physics, School of Physics, Peking University, Beijing 100871 P. R. of China
[2] CEA, IRFU, SACM, LESAR, Centre de Saclay, F-91191 Gif-sur-Yvette, France
[3] Oak Ridge National Laboratory, P.O. Box 2008, Oak Ridge, TN 37831, USA

*Abstract*

A new surface treatment technique for niobium (Nb) Superconducting Radio Frequency (SRF) cavities called Buffered Electropolishing (BEP) has been developed at JLab. It was found that BEP could produce the smoothest surface finish on Nb samples ever reported in the literature. Experimental results revealed that the Nb removal rate of BEP could reach as high as 4.09 μm/min. This is significantly faster [1] than that of the conventional electropolishing technique employing an acid mixture of HF and $H_2SO_4$. An investigation is underway to determine the optimum values for all relevant BEP parameters so that the high quality of surface finish achieved on samples can be realized within the geometry of an elliptical RF cavity. Toward this end, single cell Nb cavities are being electropolished with BEP electrolyte at both CEA-Saclay and JLAB. These cavities will be RF tested and the results will be reported through this presentation.

## INTRODUCTION

Surface polishing is one of the most important aspects of the surface treatments of Nb superconducting radio frequency (SRF) cavities to be used in particle accelerators. In the past, different techniques were applied to make the inner surfaces of SRF cavities smoother including chemical polishing, buffered chemical polishing (BCP), and electropolishing (EP). Recently, scientists at Jefferson Lab developed [2] a new Nb surface treatment technique named buffered electropolishing (BEP). BEP has been shown to be able to produce the smoothest surface finish on Nb [1,3]. Furthermore, it was also found that Nb removal rate could be as high as 4.09 μm/min. This is significantly higher than 0.3~0.4 μm/min of the conventional EP. Since it is normally required that a thickness of 150 μm of Nb has to be removed from the inner surfaces of Nb SRF cavities in order to get rid of the mechanically damaged layer as well as any evaporated Nb scale deposited on the surfaces during welding [4]. A surface treatment technique with a much higher Nb removal rate will, therefore, be beneficial to the cost of Nb SRF surface preparation and can potentially lead to capital saving in the manufacture of Nb SRF cavities to be used in particle accelerators.

Based on the experimental results from small sample studies, an effort was made to do BEP on Nb SRF single cell cavities. For this purpose, a low-cost, simple, and reliable vertical electropolishing system was set up at JLab. Cavities were also treated at CEA Saclay in France using the horizontal electropolishing system there for comparison. To understand the polishing mechanism, a demountable cavity was designed and fabricated at JLab and was shipped to Peking University for doing investigations there. Up to now, the highest accelerating gradient achieved from BEP treatments on a large grain cavity is 32 MV/m and on fine grain cavities is 23 MV/m. Preliminary results of this effort from JLab, Peking University, and CEA Saclay will be reported in this paper. A more detailed report will be published elsewhere.

## EXPERIMENTAL

The focus of this study is BEP treatment on Nb SRF single cell cavities. The demountable cavity and experiments on small flat Nb samples were employed to help understand the possible problems encountered during cavity treatments.

The acid mixture was done according to a recipe that A.T. Wu used about 15 years ago when he was working at International Superconductivity Technology Centre in Tokyo, Japan. The ratio between HF, $H_2SO_4$, and lactic acid is 4:5:11. Two different mixing orders were used: a) adding $H_2SO_4$ (96%) first to lactic acid (88%) followed by adding HF (48.8~49.2%). It is important that mixing is done with constant stirring and external cooling to keep the temperature of the electrolyte below 35 °C. b) adding HF to lactic acid first followed by adding $H_2SO_4$. This is a mixing order invented by Peking University. This mixing order is used for all the BEP work done at Peking University. Details about the acid mixing order and its possible effects on BEP will be discussed elsewhere.

Fig.1 shows the vertical electropolishing system that was constructed at JLab. Two major components of this system are a power supply for providing the potential for BEP and a refrigerated recirculator for keeping the temperature of the BEP electrolyte during polishing below 35 °C. The power supply we used was made by Electronic Measurements Inc. Model # 40T250 that could deliver a voltage as high as 40 V and current up to 250 A. Cooling of the BEP electrolyte was provided by a recirculator made by Neslab Instrument Inc. Model # CFT-150 that can deliver a cooling power of 3.75 KW in the temperature ranging from 21 to 30 °C. The operating principal of this system is schematically illustrated in Fig.2. For polishing, Pump A is turned on and Valves 2 and 4 are open. The flow rate of the electrolyte is regulated via Valve 2 with a range from 0 to 10 L/min. Valve 4 is installed to branch out the electrolyte in order to avoid labouring of Pump A. Higher flow rate is possible by opening Valve 5. Draining is achieved by opening Valves 1 and 3 and closing Valves 2, 4, and 5. For rinsing, DI water is introduced to the acid container. Then Valves 2 and 5 are open and Valves 1, 3, and 4 are closed. The cavity is flipped half way through the polishing process in order have a more homogenous polishing between upper and lower parts of the cavity.

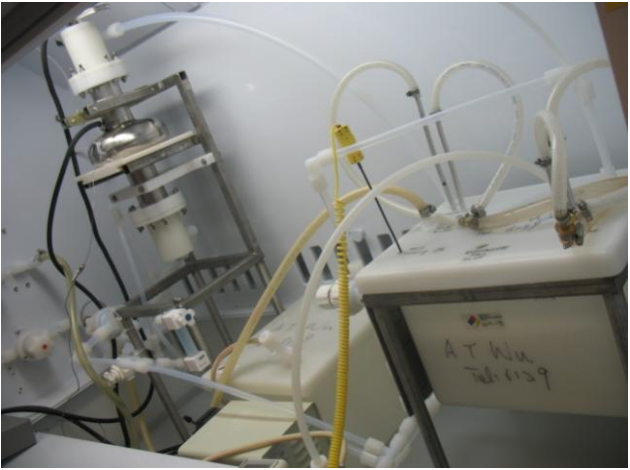

Fig.1: Vertical electropolishing system built for treating Nb SRF single cell cavities.

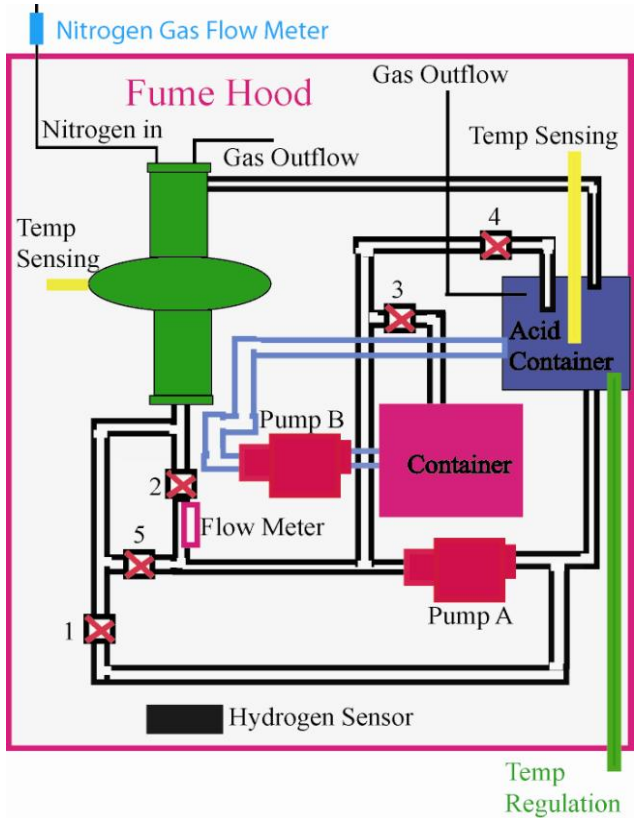

Fig.2: Schematic of the working principal of the vertical electropolishing system for Nb SRF single cell cavities (see text for more details).

Small sample experiments were carried out at Peking University, CEA Saclay, and JLab. The experimental setups were already described elsewhere [5-7]. All the small samples were cut from Nb fine grain sheets that were used for fabricating Nb SRF cavities.

## RESULTS AND DISCUSSIONS

*Effect of Electrolyte Flow Rate*

One parameter that was not studied in our previous work is the effect of electrolyte flow rate during BEP. This work was done at Peking University. Table 1 summarizes the experimental results. It is clear that Nb removal rate can be as high as 4.09 μm/min when the stirring speed is 25 Hz. Nb removal rate increases almost linearly with the stirring speed. At this moment, it is unclear what the upper limit is for Nb removal rate since the maximum stirring speed of the instrument is 25 Hz.

Although higher Nb removal rate is beneficial for reducing the cost of the fabrication of Nb SRF cavities, other factors that are known to affect the performance of the cavities have to be taken into account, including the inner surface roughness. More and more experimental data seem to indicate that a cavity with smoother inner surface finish tends to show better RF performance. To check how surface roughness is affected by electrolyte flow rate, a high precision 3-D profilometer [8] is employed. Measurements were done over an area of 200X200 μm$^2$ at minimum three locations of the surface and then took an average over the obtained root mean square (RMS). The result of RMS measurements is summarized in Table 1. Surprisingly the surfaces are not that smooth at low current densities and low Nb removal rates. The smoothest surface of RMS of is obtained on the sample having a removal rate of 3.78 μm/min. This can be attributed to the combined effects of processing time and removal rate. At a removal rate close to 3.78 µm/min, preferential attacks on grain boundaries and other defect locations are minimized.

Table 1: Summary of Experimental Results Obtained at PKU

| Sample Number | Temperature (C) | Stirring Speed (Hz) | Voltage (V) | Thickness Removed (um) | Current Density (mA/cm2) | Removal Rate (um/min) | RMS (nm) |
|---|---|---|---|---|---|---|---|
| 104 | 25 | 0 | 11.0 | 25.48 | 85.8 | 1.274 | 277.3 |
| 105 | 24 | 5 | 12.5 | 35.65 | 117.7 | 1.783 | 161.9 |
| 106 | 24 | 10 | 12.5 | 42.87 | 155.4 | 2.143 | 215 |
| 103 | 23 | 15 | 13.0 | 54.49 | 183.9 | 2.725 | 99.8 |
| 107 | 23 | 17 | 13.5 | 40.34 | 145.0 | 2.017 | 241.7 |
| 101 | 27 | 19 | 14.0 | 75.65 | 260.7 | 3.782 | 49.1 |
| 102 | 28 | 25 | 17.0 | 61.40 | 288.9 | 4.093 | 72.9 |

*BEP Treatments on Nb SRF Single Cell Cavities*

The first RF test results on a large grain Nb single cell cavity of ILC shape treated by the vertical EP system built at JLab is shown in Fig.4. After removing 75 µm from the inner surface of the cavity by BEP with an elliptical cathode shape, the inner surface looks very shiny and smooth. RF test showed a strong Q slope starting near 20 MV/m without field emission. Baking at 120 °C for 48 hrs improved the performance to 32 MV/m. This value is still lower than 35 MV/m before BEP. Since this is the very first trial of BEP on Nb single cell cavity, we consider the result encouraging.

To save time from mixing the electrolyte, we decided to order the electrolyte from a commercial supplier. First we tried to do polishing at a flow rate of 1L/min with current densities of 45, 75, and 105 mA/cm$^2$. Then repeated the same experiment at a flow rate of 7 L/min. RF test results were disappointing. Most data show a Q degradation starting near 6 MV/m without field emission. Some tests were limited by field emission due to a wrong cavity assembly procedure. All these BEP treatments were done by the commercially mixed electrolyte and the color of the commercial electrolyte looked darker than normal. Therefore we tried to do small samples experiments using the commercial electrolyte to check whether the electrolyte was normal. Test results indicated that the HF concentration of the commercial electrolyte might be significantly lower. After mixing new electrolyte in the Peking University way, using a new ball shape cathode (Fig.4), and adopting a flow rate much higher than 10L/min, 75 µm Nb removal from the inner surface of a fine grain cavity resulted in an accelerating gradient of 23 MV/m as shown in Fig.5. In this case, a much higher acid flow rate combined with the ball shape cathode helped change the flow pattern inside the cavity. Our experiment implies that a high speed electrolyte movement during BEP may be very important. Small sample studies from

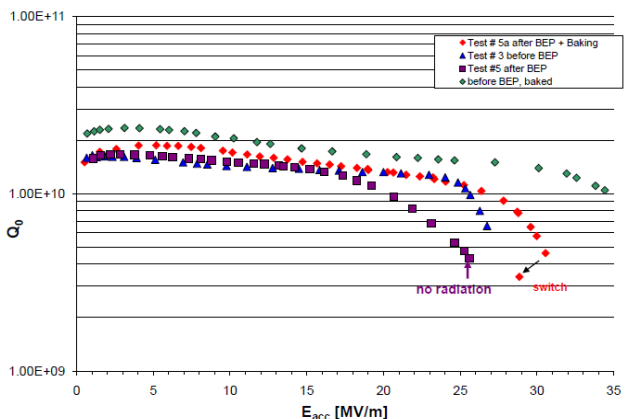

Fig.3: RF data of the first large grain Nb single cell cavity treated by the vertical EP system built at JLab.

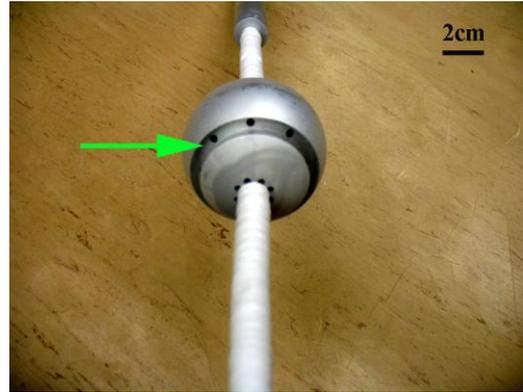

Fig.4: Ball shape cathode. The groove indicated by the arrow is designed to trap hydrogen generated during BEP.

CEA Saclay seem to indicate that the polishing mechanism of BEP is different from that of EP. Since this is a new polishing technique, more work is needed before the process can be optimized.

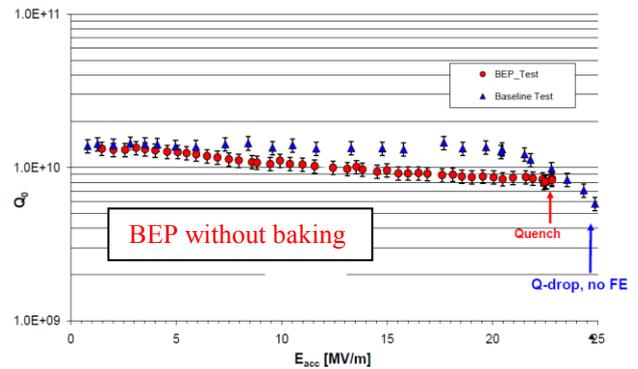

Fig.5: Excitation curves for a fine grain Nb single cell cavity treated by BEP using home-mixed acid.

## ACKNOWLEDGMENT

Thanks to P. Kneisel and A. Crawford for various help and advices during this work.